\documentclass[]{emulateapj}
\usepackage{apjfonts}

\shorttitle{HD\,271791: an extreme supernova run-away B star}
\shortauthors{Przybilla et al.}

\begin{document}

\title{HD\,271791: an extreme supernova run-away 
B star escaping from the Galaxy\footnote{Based on observations obtained at the European Southern
Observatory, proposal 073.D-0495(A).}}

\author{Norbert Przybilla\altaffilmark{2}, M. Fernanda
Nieva\altaffilmark{2,3}, Ulrich Heber\altaffilmark{2} and Keith
Butler\altaffilmark{4}}
\email{przybilla@sternwarte.uni-erlangen.de}
\altaffiltext{2}{Dr. Karl Remeis-Observatory, University of
Erlangen-Nuremberg, Sternwartstr.\,7, D-96049 Bamberg, Germany}
\altaffiltext{3}{Max-Planck-Institute for
Astrophysics, Karl-Schwarzschild-Str. 1, D-85741 Garching, Germany}
\altaffiltext{4}{University Observatory, Scheinerstr. 1, D-86179 Munich,
Germany}

\begin{abstract}
Hyper-velocity stars (HVSs) were first predicted by theory to be the result of 
the tidal disruption of a binary system by a super-massive black hole (SMBH) that 
accelerates one component to beyond the Galactic escape velocity (the Hills
mechanism). Because the Galactic centre hosts such a SMBH it is the 
suggested place of origin for HVSs. However, the SMBH
paradigm has been challenged recently by the young HVS HD\,271791 because its 
kinematics point to a birthplace in the metal-poor rim of the Galactic disc. Here we report the
atmosphere of HD\,271791 to indeed show a sub-solar iron abundance along with an
enhancement of the $\alpha$-elements, 
indicating capture of nucleosynthesis products from a supernova or a more
energetic hypernova.
This implies that HD\,271791 is the surviving secondary of a massive 
binary system disrupted in a supernova explosion.
No such run-away star has ever been found to exceed the Galactic escape
velocity, hence HD\,271791 is the first hyper-runaway star.
Such a run-away scenario is an alternative to the Hills mechanism for 
the acceleration of some HVSs with moderate velocities.
The observed chemical composition of HD\,271791 puts invaluable
observational constraints on nucleosynthesis in a 
supernova from the core-collapse of a very massive 
star ($M_{\rm ZAMS}$\,$\gtrsim$\,55\,$M_{\odot}$), which may be observed as a 
gamma-ray burst of the long-duration/soft-spectrum type.
\end{abstract}

\keywords{Galaxy: halo -- stars: abundances -- 
stars: individual (HD\,271791) -- supernovae: general}

\section {Introduction}
Studies of stellar dynamical processes in the vicinity of a super-massive 
black hole (SMBH) predicted the existence of so-called hyper-velocity 
stars (HVSs), objects moving faster than the Galactic escape 
velocity \citep{hills88,yu03}. Besides being indicators for the 
properties of the SMBH in the Galactic Centre (GC) and the surrounding stellar population, 
HVSs are also regarded as valuable
probes for the shape of the Galactic dark matter halo because of the
unique kinematic properties of stars expelled from the GC \citep{gnedin05,yu07}.

The first HVSs have only 
recently been discovered seren\-dipitously
\citep[][Paper\,I]{brown05,edelmann05,hirsch05,heber08a}.
A total of eleven faint blue stars at high Galactic latitude have been identified 
as HVSs so far after a systematic search \citep[see][]{brown07}.

All except two were found to be compatible with a scenario of ejection by the 
SMBH in the Galactic centre (GC). One exception is the early B-type 
star \object[HE 0437-5439]{HE\,0437$-$5439} which probably originates 
in the Large Magellanic Cloud (LMC, Przybilla et al.~2008), possibly expelled 
dynamically by interaction of massive binaries in a dense star cluster
\citep{leonard91} or by an intermediate-mass black hole \citep{GPZ07}, 
as no SMBH is known to exist in the LMC.

A measurement of the proper motion of the bright (V\,$=$\,12.26\,mag) early 
B-type giant \object[HD 271791]{HD\,271791} allowed the Galactic rest-frame
velocity to be determined to lie within the range 530\,--\,920\,km\,s$^{-1}$,
exceeding the local Galactic escape velocity and thus qualifying the
star as a HVS -- and the place of birth to be constrained to the outer rim
of the Galactic disk at a Galactocentric distance of $\gtrsim$15\,kpc
(Paper\,I). This left the question of the ejection mechanism to be answered
as models invoking the SMBH in the GC are ruled out.

\section {Analysis}

\begin{figure*}
\centering
\epsscale{.995}\plotone{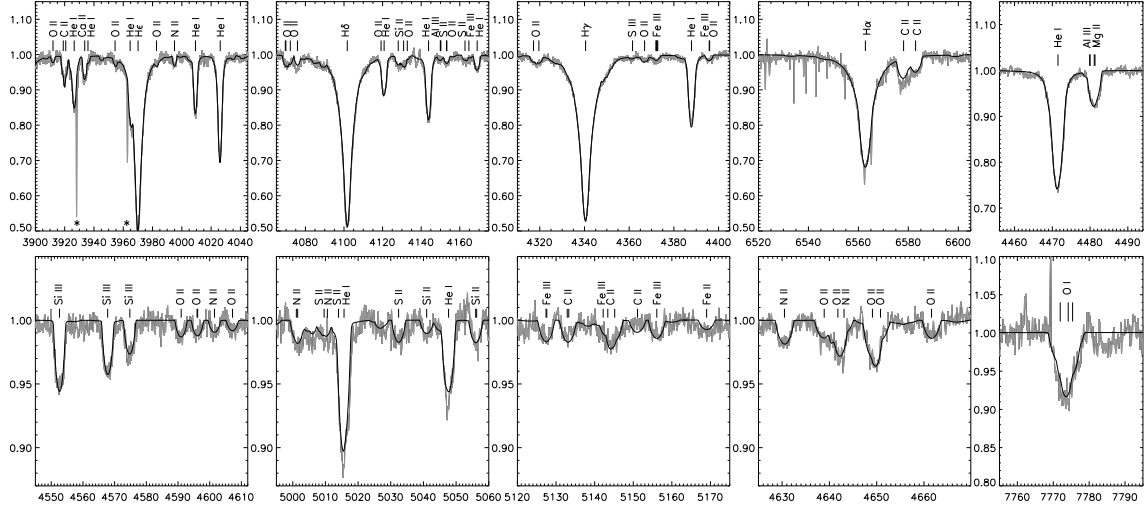} 
\caption{Comparison of spectrum synthesis for HD\,271791 (full line) with observation (grey). 
The upper panels illustrate the excellent fit of Balmer
and \ion{He}{1} lines. The ionization equilibria of \ion{Si}{2/III}, \ion{O}{1/II},
\ion{S}{2/III} and \ion{Fe}{2/III} are perfectly matched (lower panel, plus
\ion{S}{3} $\lambda$4361\,{\AA} above). A
selection of lines from most ions detected can be found in the panels. Note
that all atmospheric features of the observed spectrum are simultaneously
reproduced by the final model spectrum (for abundances as in Table~\ref{table1}). 
Abscissa is wavelength (in {\AA}), ordinate is normalized flux. 
Interstellar \ion{Ca}{2} lines are marked by an asterisk.\\
\label{fits}}
\end{figure*}

HD\,271791 was observed in early 2005 at the European Southern Observatory 
on La Silla, Chile, using the FEROS spectrograph on the 2.2\,m telescope
(Paper\,I).
Four individual spectra with resolving power $\lambda/\Delta \lambda$\,$=$\,48\,000 were
coadded and smoothed over 5 pixels for the present work, resulting in a S/N of $\sim$120 to
$\sim$160 in the blue visual range.

The quantitative analysis of the spectrum of HD\,271791 was carried 
out following the hybrid NLTE approach discussed by \citet{nieva07,nieva08}
and \citet{przybillaetal06}. 
In brief, line-blanketed LTE model atmospheres were computed with ATLAS9
and ATLAS12 \citep{kurucz93,kurucz96} -- the latter in order to
account for the chemical peculiarity of the star. NLTE line-formation
calculations were performed using updated versions of {\sc Detail} and 
{\sc Surface} \citep{giddings81, butler85}. State-of-the-art model atoms
were adopted (see Table~\ref{table1}), which allow absolute elemental
abundances to be obtained with high accuracy.

Multiple independent spectroscopic indicators are considered simultaneously 
for the determination of the atmospheric parameters, effective temperature
$T_{\rm eff}$ and (logarithmic) surface gravity $\log g$: all Stark-broadened Balmer lines
and simultaneously four metal ionization equilibria, of \ion{Si}{2/III},
\ion{O}{1/II}, \ion{S}{2/III} and \ion{Fe}{2/III}. The resulting redundancy helps 
to avoid systematic errors. The microturbulent velocity $\xi$ was determined
in the standard way by demanding abundances to be independent of line
equivalent widths. 
Elemental abundances and the (projected)
rotational velocity $v \sin i$ were determined from
fits to individual line profiles. Fundamental stellar parameters such as
mass $M$, radius $R$ and luminosity $L$, as well as the evolutionary
age $\tau_{\rm evol}$ were constrained by comparison with stellar evolution
tracks, in analogy to Paper\,I.

\begin{table}
\caption{Stellar parameters \& elemental abundances of HD\,271791 \label{table1}}
\begin{tabular}{lrr@{\hspace{12mm}}lrr}
\tableline\\[-3mm]
\tableline
$T_{\rm eff}$\,(K) & \multicolumn{2}{l}{18\,000$\pm$500} & $M/M_{\sun}$ &
\multicolumn{2}{l}{11$\pm$1}\\
$\log g$\,(cgs)    & \multicolumn{2}{l}{3.10$\pm$0.10}   & $R/R_{\sun}$ &
\multicolumn{2}{l}{15.5$\pm$0.3}\\
$\xi$\,(km/s)      & \multicolumn{2}{l}{4$\pm$1}         & $\log L/L_{\sun}$ &
\multicolumn{2}{l}{4.35$\pm$0.05}\\
$v \sin i$\,(km/s) & \multicolumn{2}{l}{124$\pm$2}       & $\tau_{\rm
evol}$\,(Myr) & \multicolumn{2}{l}{25$\pm$5}\\
\tableline
Ion & $\varepsilon^{\rm NLTE}$ & \# & Ion & $\varepsilon^{\rm NLTE}$ & \#\\
\tableline
\ion{He}{1}$^{\rm a}$ & 10.99$\pm$0.05 & 15 & \ion{Al}{3}$^{\rm h}$ & 6.33$\pm$0.08 & 3\\
\ion{C}{2}$^{\rm b}$  &  8.32$\pm$0.12 &  4 & \ion{Si}{2}$^{\rm i}$ & 7.72$\pm$0.06 & 3\\
\ion{N}{2}$^{\rm c}$  &  7.60$\pm$0.09 &  9 & \ion{Si}{3}$^{\rm i}$ & 7.71$\pm$0.10 & 5\\
\ion{O}{1}$^{\rm d}$  &  8.85$\pm$0.07 &  2 & \ion{S}{2}$^{\rm j}$  & 7.02$\pm$0.10 & 6\\
\ion{O}{2}$^{\rm e}$  &  8.81$\pm$0.07 & 14 & \ion{S}{3}$^{\rm j}$  & 7.08$\pm$0.11 & 2\\
\ion{Ne}{1}$^{\rm f}$ &  8.01$\pm$0.08 &  2 & \ion{Fe}{2}$^{\rm k}$ & 7.30$\pm$0.10 & 1\\
\ion{Mg}{2}$^{\rm g}$ &  7.28$\pm$0.12 &  1 & \ion{Fe}{3}$^{\rm l}$ & 7.22$\pm$0.08 & 6\\
\tableline
\end{tabular}\\
$\varepsilon(X)=\log\,(X/{\rm H})+12$. Statistical 1$\sigma$-uncertainties
determined from the line-to-line scatter (\# lines) are given. Estimates for
\ion{Mg}{2} and \ion{Fe}{2}.
NLTE model atoms: H: \citet{przybilla04}; $^{\rm a}$\,\citet{przybilla05};
$^{\rm b}$\,\citet{nieva06,nieva08}; $^{\rm c}$\,\citet{przybilla01}; $^{\rm
d}$\,\citet{przybillaetal00}; $^{\rm e}$\,\citet{becker88}, updated; $^{\rm
f}$\,\citet{morel08}; $^{\rm g}$\,\citet{przybillaetal01}; $^{\rm
h}$\,\citet{dufton86}; $^{\rm i}$\,\citet{becker90}, extended \& updated;
$^{\rm j}$\,\citet{vrancken96}, updated; $^{\rm k}$\,\citet{becker98};
$^{\rm l}$\,\citet{morel07}.
\end{table}

The results are summarized in Table~\ref{table1} and a comparison of the 
final model spectrum with many strategic regions of the observed spectrum 
is made in Fig.~\ref{fits}. An excellent match between the observed and 
synthetic spectrum is obtained from the large scale to minute details. 
Our stellar parameters for HD\,271791 are consistent with earlier 
work \citep[Paper\,I,][]{kilkenny88}, but more precise. This applies in particular 
to elemental abundances which have uncertainties of only about 
20\% (statistical 1$\sigma$ standard deviation). 

The uncertainties in the atmospheric parameters were constrained from the
quality of the simultaneous fits to all diagnostic indicators. Statistical
uncertainties for metal abundances were determined from the
individual line abundances. 
Systematic errors in the abundances due 
to uncertainties in atmospheric parameters, atomic data and the quality of the
spectrum are about 0.1\,dex \citep{nieva08,przybillaetal06}, i.e. as large
as the statistical errors.

\begin{figure}
\epsscale{.995}\plotone{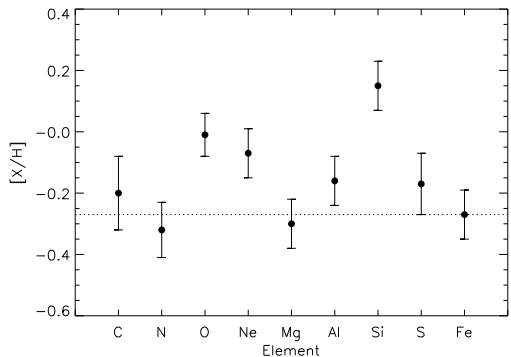}
\caption{Metal abundance pattern of HD\,271791, relative to solar
values \citep{grevesse98}. Error bars account for statistical
1$\sigma$-uncertainties. The baseline metallicity of HD\,271791, [Fe/H], 
is marked by the dotted line.
\label{pattern}}
\end{figure}

Elemental abundances of HD\,271791 (averaged over all lines of an element)
are displayed in Fig.~\ref{pattern}, relative to the solar standard. 
All abundances are within a factor of 2 relative to
solar, confirming the findings from 
Paper\,I.
However, the higher accuracy~of the present work allows an underlying pattern in the
elemental abundances to be uncovered, which is atypical for a young 
Population\,I star. The Fe abundance is subsolar by
0.27\,dex, indicating that the star was formed in a metal-poor environment. 
This is consistent with the outer metal-poor~rim~of~the Galactic disc 
suggested by the star's kinematic properties (Paper\,I). 
The $\alpha$-process elements O, Ne, S and in particular Si are enhanced, 
with the exception of Mg, which along with C and N is compatible with the 
iron underabundance by about a factor 2 relative to solar. 
This is the first detection of $\alpha$-enhancement in a massive star 
outside the GC. 

$\alpha$-enhancement is a characteristic of material ejected in the 
supernova (SN) explosion of a massive star. In-situ formation of these 
elements in HD\,271791 via hydrostatic nuclear burning and subsequent 
mixing into the stellar atmosphere can be excluded. Typically, only low-mass 
stars of the old Population\,II and the GC population show a chemical 
composition with $\alpha$-enhancement in the Galaxy at present. Membership of 
the massive B-star HD\,271791 to either population is ruled out. Note that
HD\,271791 cannot be of GC origin (where 
[Fe/H]\,$\approx$\,0 and [$\alpha$/Fe]\,$\approx$\,0.3, 
e.g. Cunha et al.~2007) by reason of its chemical 
signature in addition to its kinematics.

However, contamination of the surface layers 
of a star in a binary system is possible when the companion explodes in 
a SN, as reported e.g. for the secondary of the low-mass X-ray binary 
\object[GRO J1655-40]{Nova Scorpii 1994} \citep{israelian99}. A scenario is therefore 
developed in 
which the violent disruption of a massive binary by a SN-like event 
accounts for both, the $\alpha$-enhancement of the atmosphere of 
the surviving secondary (HD\,271791) and its acceleration to hyper-velocity.

\section{Binary scenario for accretion of SN ejecta and acceleration}
Any plausible progenitor system has to include a very massive 
primary, as the travel time of HD\,271791 from the Galactic disc 
to its present location in the halo requires the SN-like event to 
happen early in the star's lifetime. Stars above 40\,$M_{\odot}$ 
die within less than 6\,Myrs \citep{meynet05}, which is sufficiently short 
for this scenario. 

Mass accretion from a SN-shell by the secondary depends 
strongly on geometric factors because of the high velocity of the ejecta 
compared to the escape velocity from the 
surface of the secondary (several 10$^3$ vs. $\sim$10$^3$\,km\,s$^{-1}$). Interaction of the SN-shell 
with the secondary is therefore possible for a fraction of mass equal to the 
solid angle subtended from the primary at most \citep[][FA81]{fryxell81}, unless there 
is significant fall-back of material onto the remnant.
A close binary system is therefore favoured, similar to the scenario developed 
for Nova Sco 1994 \citep[][see in particular their Fig.~2]{podsi02}. 
In the present case, both components must 
be massive and the explosion of the primary has to be strongly asymmetric to 
disrupt the system and to release the secondary at its orbital velocity. 

The absence of N enrichment indicates that HD\,271791 did not accrete 
significant amounts of (CN-processed) mass from the primary, i.e. the system 
avoided Roche-lobe overflow \citep[][V98]{vanbeveren98}. This implies that the system 
has undergone a common-envelope phase, with spiral-in of the secondary because 
of viscous forces. Systems with mass ratios (secondary/primary mass)
$q$\,$\leq$\,0.2 qualify for a wide range of initial periods
(V98), resulting in a minimum mass of $\sim$55\,$M_{\odot}$ for 
the primary on the zero-age main sequence. 

The spiral-in process can expel the H-rich envelope of the primary 
(stopping the merger) when sufficient orbital energy is deposited, at the expense 
of a very large reduction of the orbital separation (V98).
Accordingly, a binary composed of a Wolf-Rayet (WR)\,$+$\,an early B-type main-sequence 
star is a conceivable pro\-geni\-tor system for the SN-like event involving 
HD\,271791. The radius of an 11\,$M_{\odot}$ star is $\sim$4\,$R_{\odot}$ 
on the main sequence, and $\sim$2-3\,$R_{\odot}$ for a WR star with mass
$\leq$\,20\,$M_{\odot}$ of WC subclass at the baseline metallicity of
HD\,271791 \citep{crowther02}. As a consequence, the separation $a$ between the binary 
components before the SN-like event could be as small as 13-15\,$R_{\odot}$ for a 
detached system, resulting in an orbital period of $\sim$1\,day and an orbital velocity 
of the secondary of $\sim$400\,km\,s$^{-1}$. 

At this point a first phase of accretion onto HD\,271791 is likely to occur. 
The WR star has a dense and strong wind that will directly impact on the 
surface of the secondary, as a metal-poor B1\,V star has no significant wind of its own 
\citep[a similar, but more extreme situation to that observed for the massive binary
CPD\,$-$41\degr\,7742,][]
{sana05}. Using conservative assumptions on WR mass-loss rates and lifetime, 
we estimate that about 0.04\,$M_{\odot}$ of material,
rich in C and to a lesser degree abundant in O and Ne, could be deposited.

The final collapse of the WC star into a black hole \citep{heger03} results in an 
ordinary type Ic\,SN or 
a more energetic hypernova (HN). The kick experienced by the (proto-) black hole disrupts 
the binary and releases the companion at its orbital velocity, in analogy to the classical 
scenario of run-away stars \citep{blaauw61}. 

This gives rise to the second -- and main -- mass accretion event onto HD\,271791.
The interaction of the expanding SN-shell with the secondary is a 
complex hydrodynamical process involving ablation of mass from the outer 
layers of the secondary and accretion from the shell, which is not completely 
understood (FA81). Also interaction with fall-back material from the 
explosion can be significant \citep{podsi02}. We therefore made conservative assumptions 
to estimate the effects of accretion of SN-processed material on the chemical 
composition of the secondary. Nucleosynthesis yields of \citet{nomoto06} were used.
We chose the highest mass models available, for a star of initially
40\,$M_{\odot}$ and metallicity $Z$\,$=$\,0.004,
that comes closest to the adopted pristine value. The total kinetic energy
of the explosion predicted by the models is 3$\times$10$^{52}$\,erg for the HN and 
10$^{51}$\,erg for the SN, respectively. A total of $M_{\rm
acc}$\,$=$\,$M_{\rm ejecta}\,R_2^2/4a^2$
may be accreted by the secondary with radius $R_2$.
However, due to the large momentum of the impacting material the efficiency of accretion 
is small, i.e. $\sim$10\% (FA81). 
About 0.02\,$M_{\odot}$ of heavy elements can therefore be accreted 
in the event that expels $M_{\rm ejecta}$\,$\approx$\,10\,$M_{\odot}$ of
metals in total (for $a$\,$=$\,13 to 15\,$R_{\odot}$).
At the same time $\sim$1\% (0.1\,$M_{\odot}$) of the secondary 
mass is ablated from the surface layers (FA81), including a considerable part 
of the material accreted previously from the WR wind.

The accreted material has been mixed with unpolluted matter from
the secondary's deeper layers and possibly with CN-processed from its core
over the past $\sim$20\,Myrs. Chemical mixing in the radiative envelopes
of rotating stars is slow, approximated well by diffusion 
\citep{maeder00}. The metal-rich material will therefore be only partially mixed, 
leading to a dilution of the original HN/SN abundance pattern. 

In order to model the observed abundance pattern, several parameters
have to be adjusted: {\sc i}) the initial chemical composition; {\sc ii}) the fraction
of the SN-shell that is accreted and {\sc iii}) the equivalent envelope mass
affected by complete mixing (in reality the heavy element
abundances will be stratified). The scenario has to reproduce the observed Fe abundance as a
boundary condition as the abundance patterns are normalised with respect to iron.
The model abundance patterns do not account for the surviving 
fraction of the C-rich and N-free WC wind and rotational mixing with slightly C-depleted 
and N-enriched material from the core as quantitative estimates are difficult to make. 
However, both processes tend to cancel each other, so that the net effect is
expected to be small.

\begin{figure}
\epsscale{.995}\plotone{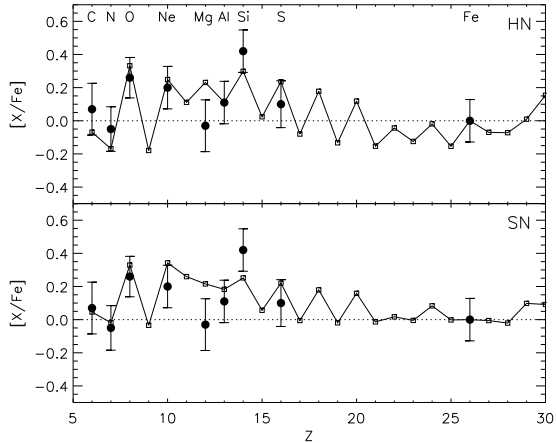}
\figcaption{Elemental abundance pattern of HD\,271791 (dots, relative to solar 
as a function of atomic number $Z$, normalized to Fe)
compared to HN/SN yields of \citet{nomoto06}. Error bars account for statistical 
and sys\-tematic uncertainties of 0.1\,dex, summed in
quadrature. See text for~details.
\label{yields}}
\end{figure}

Results are shown in Fig.~\ref{yields}, for the
mixing of 0.12\%/0.08\% of the HN/SN ejecta with 1\,$M_{\odot}$ envelope
material of HD\,271791 
of pristine composition [$X$/H]\,$=$\,$-$0.4/$-$0.24\,dex.
Note that mixing of a higher fraction of envelope mass can compensate for an increase of the
accreted mass and vice versa. 
Overall, reasonable agreement between the model predictions and observation is
obtained, except that too large a Mg and too low a Si
abundance is predicted. However, this is likely an artifact of the use of
integrated yields. More realistic is a latitudinal dependency of the ejecta
composition in the likely case of an aspherical explosion \citep{maeda08}
and a stratified abundance distribution in the ejecta.
Hydrodynamic instabilities at the trailing edge of the
SN-shell will preferentially lead to an accretion of material from the
inner shell on the secondary (FA81), which is rich in Si and depleted in Mg
\citep[e.g.][]{maeda02}. Tailored nucleosynthesis calculations for this
particular case, a detailed hydrodynamical simulation of the
SN-shell--secondary interaction and a comprehensive modeling of the
subsequent mixing of the accreted material with the secondary's envelope are
desirable.

\section{Conclusions}
An accurate determination of the surface composition of the HVS HD\,271791
in the present work confirmed the low [Fe/H] expected for a star born in the outer
Galactic rim (Paper\,I). The finding of $\alpha$-enhancement allowed the
acceleration mechanism to be identified as an extreme case of the run-away
scenario \citep{blaauw61}. 
The evolution of a binary with a very
massive primary ($M$\,$\gtrsim$55\,$M_{\odot}$) can lead to a suitable 
progenitor system of a close WR\,$+$\,B main sequence star that can account
for both, the observed $\alpha$-enhancement and ejection at high orbital
velocity. The collapse of the WR star into a black hole gave rise to either
an (aspherical) SN or a more energetic HN \citep[which may be accompanied by
a gamma-ray burst
of the long-duration/soft-spectrum type,][]{woosley06}, in which the system
was disrupted. A simple model for accretion from the SN-shell and subsequent mixing with 
pristine envelope material can qualitatively explain the observed
abundance pattern of HD\,271791. Only one fact remains to be explained:
the difference between the kick velocity from the binary disruption
($v_{\rm kick}$\,$\approx$\,400\,km\,s$^{-1}$) and the Galactic rest frame velocity of HD\,271791
($v_{\rm grf}$\,$=$\,530--920\,km\,s$^{-1}$).

Inspection of Fig.~4 in Paper\,I shows that HD\,271791 was ejected from the
Milky Way in the direction of the Galactic rotation. A rough alignment of
the orbital velocity vector of the secondary with the Galactic rotation direction 
at the moment of system disruption is therefore sufficient to explain the 
space motion of HD\,271791 if it is on the lower end of the observed range. 
Numerical kinematic experiments using the
Galactic potential of \citet{allen91} imply a minimum $v_{\rm
kick}$\,$\approx$\,380\,km\,s$^{-1}$ to obtain the minimum $v_{\rm grf}$
of HD\,271791 ($\sim$530\,km\,s$^{-1}$) that is consistent with the observed 
proper motion, radial velocity, distance and age. For the `best' proper motion of Paper\,I
($v_{\rm grf}$\,$=$\,630\,km\,s$^{-1}$) a $v_{\rm
kick}$\,$\approx$\,460\,km\,s$^{-1}$ would be required.
We conclude that the proposed SN
run-away scenario can reproduce all observational constraints, the
kinematics and the abundance pattern. Extreme cases of the run-away scenario
of \citet{blaauw61} therefore pose an alternative to the Hills mechanism for
the acceleration of some HVSs with moderate velocities.
As HD\,271791 is escaping from the Galaxy, it could be termed a
hyper-runaway star, the first of its class.

\acknowledgements
We thank R. Napiwotzki for valuable discussions on the kinematics of
HD\,271791 and the ESO staff at La Silla for support with the observations.
M.~F.~N. gratefully acknowledges financial support by the Deutsche
Forschungsgemeinschaft (grant HE 1356/45-1). 


\end{document}